\documentclass[letterpaper, 10 pt, conference]{ieeeconf}  
                                                    
\IEEEoverridecommandlockouts                              
\usepackage{graphicx}
\usepackage{caption}
\usepackage{subcaption}
\usepackage{color}
\usepackage{placeins}

\usepackage{breakurl}
\usepackage[breaklinks]{hyperref}
\begin{document}
\color{red}
\onecolumn
\huge 
\begin{center}
2019 10th IFIP International Conference on New Technologies, Mobility and Security (NTMS)
\end{center}
\vskip 1.5in
\Large 
978-1-7281-1542-2/9/ ©2019 IEEE

DOI 10.1109/NTMS.2019.8763809
\vskip 1.5in
\Large 
"\textcopyright \textcopyright 2019 IEEE. Personal use of this material is permitted. Permission from IEEE must be obtained for all other uses, in any current or future media, including reprinting/republishing this material for advertising or promotional purposes, creating new collective works, for resale or redistribution to servers or lists, or reuse of any copyrighted component of this work in other works."

\color{black}
\twocolumn

\title{\LARGE \bf
Kriptosare.gen, a dockerized Bitcoin testbed: analysis of server performance
}

\author{
\authorblockN{Francesco Zola\authorrefmark{1}, Cristina P\'erez-Sol\`a\authorrefmark{2}, Jon Ega\~na Zubia\authorrefmark{1}, Maria Eguimendia\authorrefmark{1}, Jordi Herrera-Joancomart\'{\i}\authorrefmark{3}\\ \\}
\authorblockA{\authorrefmark{1} Dept. Data Intelligence for Energy and Industrial Processes, Vicomtech, \\Paseo Mikeletegi 57, 20009 Donostia/San Sebastian, Spain\\ \{fzola, jegana, meguimendia\}@vicomtech.org}
\authorblockA{\authorrefmark{2}Internet Interdisciplinary Institute (IN3), Universitat Oberta de Catalunya (UOC),\\CYBERCAT-Center for Cybersecurity Research of Catalonia, 08860 Castelldefels, Barcelona, Spain\\cperezsola@uoc.edu}
\authorblockA{\authorrefmark{3}Dept. d'Enginyeria de la Informaci\'o i les Comunicacions, Universitat Aut\`onoma de Barcelona,\\CYBERCAT-Center for Cybersecurity Research of Catalonia, 08193 Bellaterra, Catalonia, Spain\\jordi.herrera@uab.cat}
}

\maketitle
\thispagestyle{empty}
\pagestyle{empty}

\begin{abstract}

Bitcoin is a peer-to-peer distributed cryptocurrency system, that keeps all transaction history in a public ledger known as blockchain. The Bitcoin network is implicitly pseudoanonymous and its nodes are controlled by independent entities making network analysis difficult.
This calls for the development of a fully controlled testing environment.

This paper presents \textit{Kriptosare.gen}, a dockerized automatized Bitcoin testbed, for deploying full-scale custom Bitcoin networks. The testbed is deployed in a single machine executing four different experiments, each one with different network configuration. We perform a cost analysis to investigate how the resources are related with network parameters and provide experimental data quantifying the amount of computational resources needed to run the different types of simulations. Obtained results demonstrate that it is possible to run the testbed with a configuration similar to a real Bitcoin system.

\end{abstract}

\IEEEoverridecommandlockouts
\begin{keywords}
Bitcoin testbed, blockchain , dockerized network, server performance, regtest
\end{keywords}

\section{INTRODUCTION}
Bitcoin uses a P2P network to propagate all the information within the system\cite{nakamoto2008bitcoin}. Peers broadcast their transactions to other peers through this network and miners use it to propagate their newly found blocks.

Bitcoin is implemented using a decentralized peer-to-peer network and, because of that, there is no central point of failure where the network can be attacked.
However, it also hinders the analysis of the network. Bitcoin nodes only have a partial local view of the network, so there is no direct way to analyze the network as a whole. 

Since nodes are controlled by independent entities, there is no way to know their exact behavior. Although most nodes run standard versions of the Bitcoin Core client\footnote{\url{https://bitnodes.earn.com/nodes/}}, many also run modified versions, either for research or for ideological purposes. In any case, the exact responses and behaviour of a node with respect to certain network messages and in specific situations have been used to attack the network in the past, for instance, by fingerprinting users~\cite{biryukov2015bitcoin} or by revealing the network topology~\cite{miller2015discovering, delgado2018txprobe}. 
Not being able to see the whole network as well as not knowing how nodes behave, makes analysis of the Bitcoin network very difficult. 

The contribution of this paper is to present \textit{Kriptosare.gen}, a platform for deploying full-scale Bitcoin networks, together with the tools to control the network. \textit{Kriptosare.gen} provides automatized features that allow users to deploy a large number of Bitcoin nodes and simulate the Bitcoin node interactions in an easy way. This automatic process increases the scalability of the testbed. 

Is also important to understand under which conditions is possible to execute a simulation within \textit{Kriptosare.gen}. For this reason, the paper provides data quantifying the amount of computational resources needed to run different simulations. These results allow researchers to estimate the cost of running a simulation, and provide bounds on the simulations that can be executed in a given hardware.

With a tool that is able to reproduce the Bitcoin network in a fully controlled environment where all peers are monitored, and where data from all of them can be incorporated into the analysis, researchers and developers are able, for instance, to study how changes of the protocol affect the network or to simulate network attacks or user behaviour. 

The rest of the paper is organized as follows. Section~\ref{sec:related} describes the related work. After that, Section~\ref{sec:arch} presents the architecture of the designed testbed. Then, Section~\ref{sec:evaluation_one} provides experimental evaluation of the testbed. Finally, Section~\ref{sec:conclusions} concludes the paper and provides guidelines for further work.

\section{STATE OF THE ART}
\label{sec:related}
Simulation allows to create environments to experiment with Bitcoin networks without actually deploying them.

Simulators provide a good approach to recreate a real environment since they are often developed and optimized in order to satisfy a simulation goal. As a result, required hardware resources can be bounded, and large simulations can be performed. However, since simulators are specialized developed software (normally derived from the software that runs the real system) they need to evolve in parallel to the real software they simulate. Such a task becomes very time demanding in highly dynamic software, like Bitcoin, as it is pointed out in
\cite{Neudecker2019_1000089033}. To overcome that problem, another approach is to set an emulation environment using the standard software that the original system is using. To that end, bitcoind (the reference Bitcoin client, that implements the full Bitcoin protocol) has a regression test mode or \texttt{regtest} that allows to create private local Bitcoin networks for testing and development purposes. A bitcoind client running in \texttt{regtest} mode will follow the same rules than a \texttt{testnet} node, but will work on a new private blockchain (with a new genesis block). The main property this mode offers is the ability to arbitrarily create blocks, that is, blocks are created on demand, by issuing a special remote procedure call (RPC) command to the node, allowing developers to control the network. However, the \texttt{regtest} mode just provides a way for a Bitcoin client to work on a private network with a private blockchain, but does not offer any tools to deploy such a network, generate traffic and replicate common behaviors or monitorize and visualize its status. 

Arthur Gervais et al. created a Bitcoin simulator \cite{gervais2016security} for ns3 \cite{riley2010ns} (a popular network simulation) focused on reproducing block propagation through the network. Since they were studying block propagation, the simulator is just limited to blocks, and does not reproduce any other Bitcoin network events (propagation of transactions, network messages, etc.). A. Miller and R. Jansen implemented a Bitcoin simulator~\cite{miller2015shadow} for Shadow, that allows to run real applications over a simulated Internet topology.

There exist some basic private Bitcoin network generation platforms based on Docker~\cite{joem,freewil,cryptochain}. Nevertheless, many of them are very similar to each other in terms of implementation, lack a clear explanation, and are based on just few nodes (usually 2). Although these projects present a procedure to automatize the network creation, they do not consider the network topology nor scalability problems. Other basic private Bitcoin network generations \cite{freekpaans,ariejan} do not provide automatization for user simulation, which means that the operations of the single Bitcoin user have to be executed manually. Therefore, they require a lot of operations/effort to create a large number of nodes and to replicate user interactions within the Bitcoin network. To the best of our knowledge, these projects do not offer enough information (like needed resources) to execute a large number of nodes nor tools to monitor the network's state.

This paper fills in this gap by presenting a complete automatized Bitcoin network testbed platform that allows deployment, interaction, and resource monitoring of private Bitcoin networks. \textit{Kriptosare.gen} allows researchers to deploy a full-scale Bitcoin network, control the topology of that network, and automatically simulate user behaviour. 

\section{ARCHITECTURE}
\label{sec:arch}

\subsection{Deploying the network}

The \textit{Kriptosare.gen} testbed consists of a private Bitcoin network that replicates the real architecture on-scale. Its deployment is based on Docker, a platform that allows creating "containers" with all the required software. Containerized software will always run the same code, regardless of the environment, ensuring high portability and easy setup.
The testbed creates containers as many as there are bitcoin users, so that each container represents a real user/wallet. In its basic setting, each node runs a Bitcoin Core client instance in \texttt{regtest} mode representing a wallet in the real world.

Bitcoin clients inside each of the docker containers are controlled by the testbed through remote procedure calls sent to their \texttt{18332} port.
Additionally, each bitcoind client in each container uses port \texttt{18444} to establish connections with other nodes in the P2P network. The \texttt{regtest} mode does not include a peer discovery mechanism, so nodes do not automatically try to connect to other nodes. Therefore, connections between nodes in the testbed are created by explicitly sending RPC commands to nodes, and thus \textit{Kriptosare.gen} is able to control the topology of the network.


RPC calls to each bitcoind instance are also used to generate activity in the testbed. 

On the one hand, in order to generate transactions, the testbed creates a thread, that periodically sends transaction generation RPC commands to nodes in the testbed. Specifically, source and destination nodes are chosen randomly, and the amount to transfer is also selected randomly from the balance available in the source node's wallet. A new address is then requested to the destination node, and a transaction is sent to this newly generated address.

On the other hand, blocks are generated by randomly selecting one of the bitcoind instances as the next miner, and sending it the proper RPC call.

\subsection{Resources monitoring and storage}
\label{ssec:metrics}
In order to evaluate the resources being used by \textit{Kriptosare.gen} in realtime, we consider five different metrics. These metrics are collected by Telegraf, an agent for collecting and reporting metrics and events. 

The analyzed metrics collected from Telegraf to evaluate the performance of our testbed are:
\begin{itemize}
\item \textit{CPU usage percentage}: the percentage of CPU used by the testbed.
\item \textit{Memory usage}: the amount of RAM memory used by the testbed (GB).
\item \textit{Disk usage percentage}: the percentage of disk space needed to run the simulation.
\item \textit{Disk I/O}: the speed of the host server during the operation on a physical disk (read/write) (MB/s).
\item \textit{Network speed}: speed of network traffic within the whole testbed (KB/s). 
\end{itemize}

The first three metrics are computed directly by Telegraf. The last two are calculated from the amount of data write/read and sent/received (respectively, for Disk I/O and for Network speed), deriving the samples provided by Telegraf over time.

Data from the mentioned metrics are stored to an InfluxDB\footnote{https://www.influxdata.com/}, and then processed and visualized in Python, inside a Jupyter Notebook. InfluxDB is an open-source database suited for time series (TSDB). Once the database is created, the data are stored in data points, each of them made by a measurement, a tagset, a fieldset, and a timestamp. Data from InfluxDB are then read from a python Jupyter Notebook, and processed to extract all the information from the time range of the simulations. Batches of simulations are processed to obtain the averages and standard deviations of each of the selected metrics.

\section{Performance evaluation}
\label{sec:evaluation_one}

The amount of resources needed to run a simulation with \textit{Kriptosare.gen} depends on the parameters of the simulation one wants to run. In this section, we study the impact each of the parameters of the simulation has on the required server resources when deploying the testbed on a single machine. 

In order to evaluate the performance of the testbed when it is executed on a single machine, we run a set of \textit{simulations}. Each simulation is defined by its \textit{configuration}, that is, a set of \textit{parameters} together with the values they take in the simulation. The word \textit{batch} is used to indicate a set of simulations with the same configuration. Simulations are repeated five times with the same configuration to ensure that reported metrics are representative enough. Then, the resources needed to run the simulations are evaluated with respect to a set of \textit{metrics} (as detailed in Section~\ref{ssec:metrics}). For each metric, we provide average and standard deviations of all the simulations in a batch.

The parameters used in the configurations are: the number of Bitcoin nodes, the number of peers each node has within the Bitcoin network, the speed at which transactions are generated in the network (TX speed), and the speed at which blocks are mined in the network (BLK speed).

We define an \textit{experiment} as a set of simulations that tries to evaluate the impact of a given parameter in the resources needed to run the testbed. An experiment has three of the four parameters of the configuration fixed to a constant value, and the fourth varied within an interval. This allows us to evaluate the impact of the varied parameter on the resources needed to run the simulation. 

All the simulations are made in a virtual machine with 4 CPUs Intel(R) Xeon(R) Silver 4114 CPU @ 2.20GHz, 32GB RAM DDR4 memory with 2,666 MHz, and 100GB of Hard Disk SATA.

Table~\ref{tab:simsv1} presents a summary of the configurations of the experiments, where TX speed is measured in transactions per second and BLK speed in blocks per hour. Experiments have been created in an increasing order of complexity, starting with a basic configuration where no peers, transactions or blocks are created, and finishing with a realistic configuration where nodes have peers, generate transactions, and mine blocks.

\begin{table}[]
\setlength{\tabcolsep}{4pt}
{
\begin{tabular}{c|cccc}
Exp. Id & \# of nodes & \# connection  & TX speed  & BLK speed \\
&  & per node & gen. & gen. \\ \hline
1       & [0, 400]       & 0                  & 0                 & 0         \\
2       & 100         & [2, 16]            & 0                 & 0         \\
3       & 100         & 8                  & [0, 15]            & 0         \\
4       & 100         & 8                  & 7                 & [4, 20]    \\
\end{tabular}}
\caption{Summary of the configuration of each experiment}
\label{tab:simsv1}
\end{table}

\FloatBarrier
\subsection{Experiment 1: Number of nodes}

In the first experiment, we evaluate the impact of increasing the number of Bitcoin nodes in the network. Therefore, the number of Bitcoin nodes in the simulation is varied, starting from 0 up to 400. Nodes in this experiment do not have any connections to other peers of the network, nor generate any transactions nor blocks. Therefore, this experiment allows to account for the resources needed to create docker containers with bitcoind nodes.

\begin{figure}[!htbp]
  \centering
  \begin{subfigure}[b]{0.4\linewidth}
    \includegraphics[width=\linewidth]{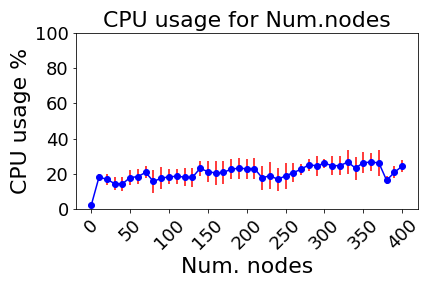}
    \caption{CPU usage}
    \label{fig:Sim1_CPU}
  \end{subfigure}
  \begin{subfigure}[b]{0.4\linewidth}
    \includegraphics[width=\linewidth]{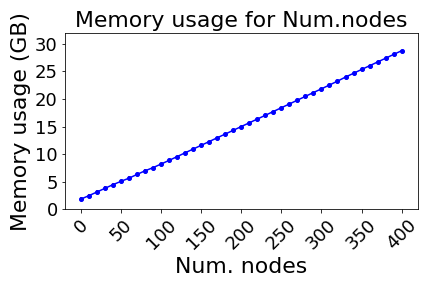}
    \caption{Memory usage}
    \label{fig:Sim1_Mem}
  \end{subfigure}
  \begin{subfigure}[b]{0.4\linewidth}
    \includegraphics[width=\linewidth]{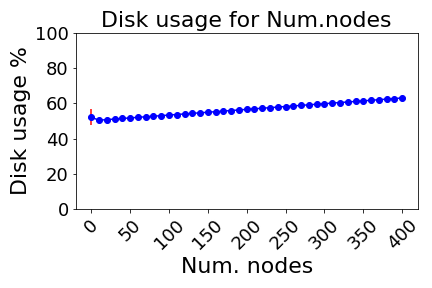}
    \caption{Disk usage}
    \label{fig:Sim1_Disk}
  \end{subfigure}
  \begin{subfigure}[b]{0.4\linewidth}
    \includegraphics[width=\linewidth]{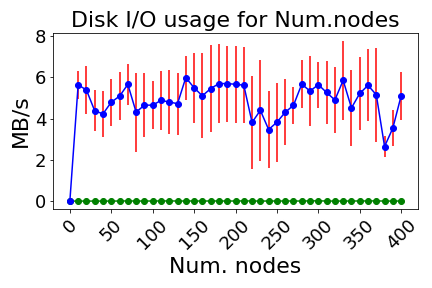}
    \caption{Disk I/O}
    \label{fig:Sim1_DIO}
  \end{subfigure}
  \caption{First experiment results: increasing the number of nodes}
  \label{fig:Sim1}
\end{figure}

Each point in the figures of this section shows the average and standard deviation of the metric as reported in a batch of five simulations with the same configuration. 

This experiment uses around $20\%$ of CPU, with an slow increase with the number of nodes (see Figure~\ref{fig:Sim1_CPU}). The consumption of RAM memory increases linearly with respect to the number of bitcoin nodes in the simulation (Figure~\ref{fig:Sim1_Mem}). $50\%$ of disk space is used by the testbed platform (without any deployed nodes), and disk usage increases with each new container deployed (Figure~\ref{fig:Sim1_Disk}). 

Up to 400 nodes can be created with the hardware resources allocated for the experiment; the amount of RAM memory used being the limiting factor.

\FloatBarrier
\subsection{Experiment 2: Number of connections per node}

In the second experiment, we evaluate the resources needed to run the simulations when there are Bitcoin nodes and when those nodes are connected to each other. Therefore, the testbed now needs to maintain the connections each node has, and manage and process the network traffic that is sent through these connections. 

All the simulations in this experiment have the number of nodes fixed to 100. The number of peers per node, ranges between 2 and 16 (tested values are $\{2, 4, 8, 12, 16\}$). The network topology for simulations in this experiment is fixed, with each peer in the network having exactly the same number of connections. 

\begin{figure}[!htbp]
  \centering
  \begin{subfigure}[b]{0.4\linewidth}
    \includegraphics[width=\linewidth]{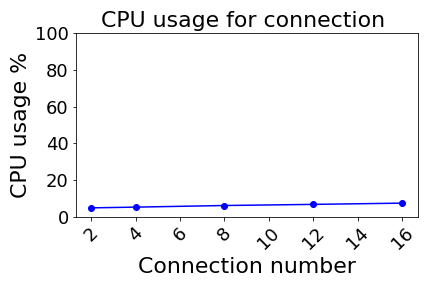}
    \caption{CPU usage}
    \label{fig:Sim3_CPU}
  \end{subfigure}
  \begin{subfigure}[b]{0.4\linewidth}
    \includegraphics[width=\linewidth]{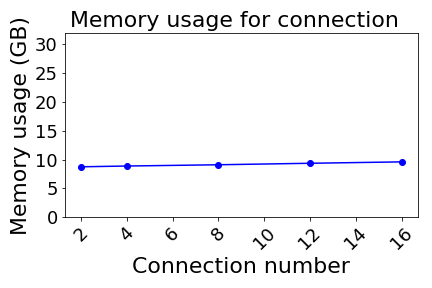}
    \caption{Memory usage}
    \label{fig:Sim3_Mem}
  \end{subfigure}
  \begin{subfigure}[b]{0.4\linewidth}
    \includegraphics[width=\linewidth]{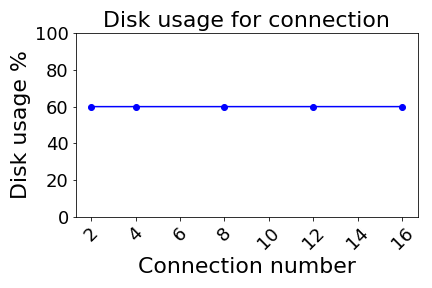}
    \caption{Disk usage}
    \label{fig:Sim3_Disk}
  \end{subfigure}
  \begin{subfigure}[b]{0.4\linewidth}
    \includegraphics[width=\linewidth]{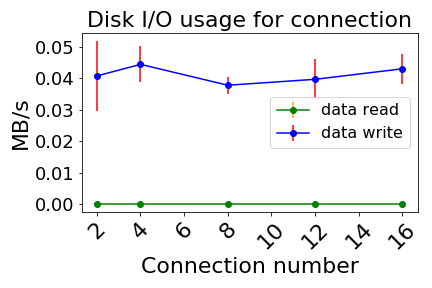}
    \caption{Disk I/O}
     \label{fig:Sim3_DIO}
  \end{subfigure}
  \begin{subfigure}[b]{0.4\linewidth}
   \includegraphics[width=\linewidth]{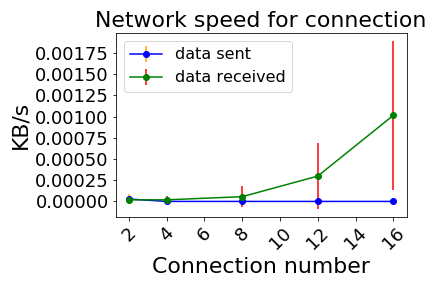}
    \caption{Network speed}
     \label{fig:Sim3_NetS}
  \end{subfigure}
  \caption{Second experiment results: increasing the number of peers per node}
  \label{fig:Sim3}
\end{figure}

Changing the number of connection between the nodes in the network has little impact on the CPU (Figure~\ref{fig:Sim3_CPU}), memory (Figure~\ref{fig:Sim3_Mem}), and disk usage (Figure~\ref{fig:Sim3_Disk}).

Note that the metrics shown in the plots of this experiment are evaluated after the containers have been created. Therefore, they account only for the resources needed to create and maintain the network connections (and maintain the docker containers), but do not take into account the cost of creating these containers. This is why CPU and memory usage reported in this experiment is lower than in Experiment 1. We decided not to consider container generation here in order to evaluate the impact of increasing the number of connections in the resources needed to run the simulation.

Increasing the number of connections per node increases the speed of network traffic in the simulation (Figure~\ref{fig:Sim3_NetS}). However, this increase is really small. The reason is that these connections are barely used, since there are no transactions nor blocks to propagate. 
This is also why increasing the number of peers per node has also minimal impact on the memory needed to run the simulation (Figure~\ref{fig:Sim3_Mem}): since no transactions nor blocks are propagated, there is no need to perform costly cryptographic operations such as hashes, signature generations, and signature validations.

\FloatBarrier
\subsection{Experiment 3: Transaction generation speed}

In the third experiment, we evaluate the performance of the testbed when transactions are generated and propagated through the network at different speeds.

In this batch, the simulations have the number of the nodes fixed to 100 and the number of connections fixed to 8. During this experiment, no blocks are yet created, so all transactions remain in the mempool of the nodes. 
The transaction generation speed ranges between $0$ and $15$ tx/s (tested values are $\{0, 1, 4, 7, 15\}$). Like in the previous experiment we do not consider the resources needed to create the containers.

\begin{figure}[!htbp]
  \centering
  \begin{subfigure}[b]{0.4\linewidth}
    \includegraphics[width=\linewidth]{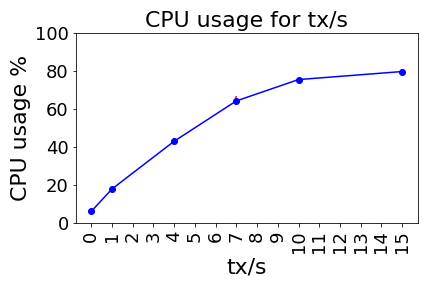}
    \caption{CPU usage}
    \label{fig:Sim2_CPU}
  \end{subfigure}
  \begin{subfigure}[b]{0.4\linewidth}
    \includegraphics[width=\linewidth]{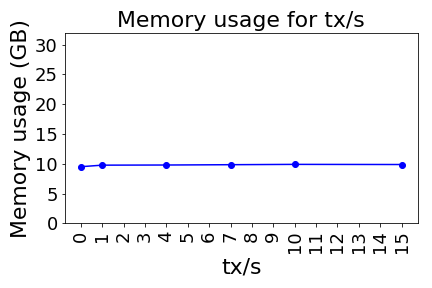}
    \caption{Memory usage}
    \label{fig:Sim2_Mem}
  \end{subfigure}
  \begin{subfigure}[b]{0.4\linewidth}
    \includegraphics[width=\linewidth]{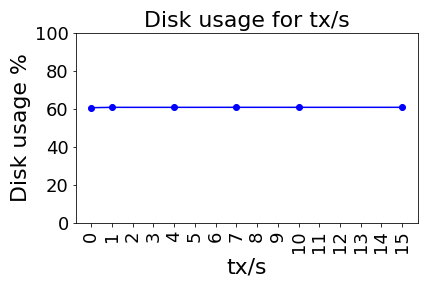}
    \caption{Disk usage}
    \label{fig:Sim2_Disk}
  \end{subfigure}
    \begin{subfigure}[b]{0.4\linewidth}
    \includegraphics[width=\linewidth]{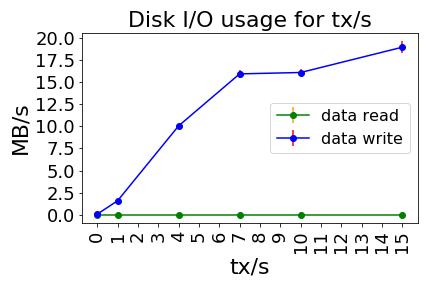}
    \caption{Disk I/O}
    \label{fig:Sim2_DIO}
  \end{subfigure}
  \begin{subfigure}[b]{0.4\linewidth}
    \includegraphics[width=\linewidth]{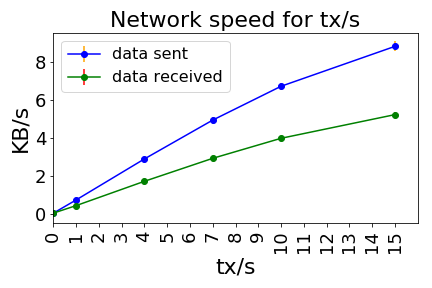}
    \caption{Network speed}
     \label{fig:Sim2_NetS}
  \end{subfigure}
  \begin{subfigure}[b]{0.4\linewidth}
    \centering\includegraphics[width=0.65\linewidth]{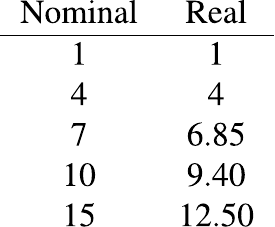}
    \caption{Tx speed (tx/s)}
     \label{tab:simnvrs}
  \end{subfigure}
  \caption{Third experiment results: increasing transaction speed}
  \label{fig:Sim2}
\end{figure}

Disk and memory usage are not affected by transaction generation speed (see Figs.~\ref{fig:Sim2_Mem} and \ref{fig:Sim2_Disk}). On the contrary, CPU usage shows a nearly lineal increase with the increment on tx generation speed. The change of tendency shown in the highest speeds is due to the procedure used to generate transactions. In our testbed, increasing tx speed implies increasing the number of threads that manage tx generation and, in turn, increases the execution time of a single RPC call. This behaviour slows down the tx speed, creating a small difference between the nominal value (the speed chosen a priori) and the real value (as shown in Figure~\ref{tab:simnvrs}).

Increasing the tx speed increases CPU usage because generating transactions and propagating them bears computational effort in terms of cryptographic operations. Transactions in the simulation spend a single P2PKH output and create a single P2PKH output, and thus one signature has to be created for each generated transaction, and one signature validated for each received transaction.

Additionally, the hash of the transaction is also computed in order to be able to generate and validate the signature. Increasing transaction generation speed also increases the network speed (Figure~\ref{fig:Sim2_NetS}). 

\FloatBarrier
\subsection{Experiment 4: Block generation speed}

In the last experiment, we evaluate the performance of the server with respect to block generation speed. The number of nodes is $100$, the number of connections per peer is $8$, the transaction generation speed is $7$ tx/s, and block generation speed is variable, ranging from $4$ to $20$ blocks/h (tested values are $\{4, 8, 12, 16, 20\}$). Like the previous two experiments we do not consider the resources needed to create the containers.

\begin{figure}[!htbp]
  \centering
  \begin{subfigure}[b]{0.4\linewidth}
   \includegraphics[width=\linewidth]{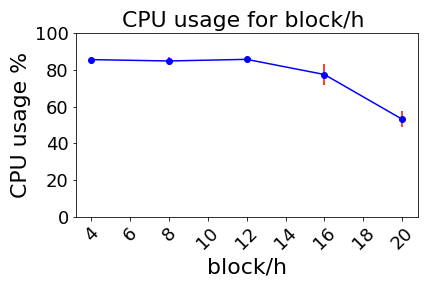}
    \caption{CPU usage}
    \label{fig:Sim4_CPU}
  \end{subfigure}
  \begin{subfigure}[b]{0.4\linewidth}
    \includegraphics[width=\linewidth]{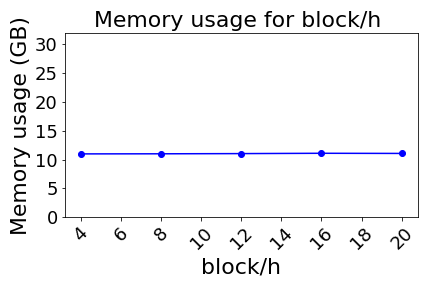}
    \caption{Memory usage}
    \label{fig:Sim4_Mem}
  \end{subfigure}
  \begin{subfigure}[b]{0.4\linewidth}
    \includegraphics[width=\linewidth]{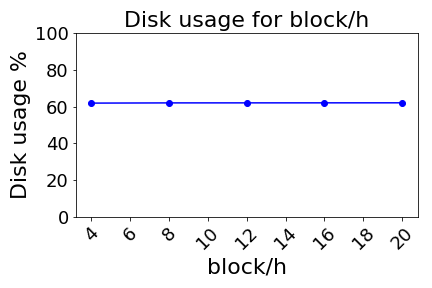}
    \caption{Disk usage}
    \label{fig:Sim4_Disk}
  \end{subfigure}
  \begin{subfigure}[b]{0.4\linewidth}
   \includegraphics[width=\linewidth]{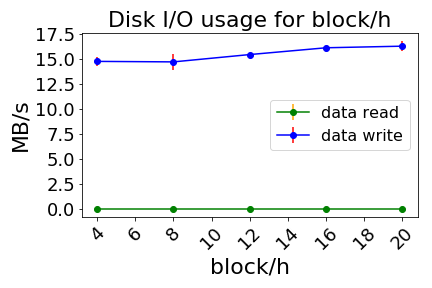}
    \caption{Disk I/O}
    \label{fig:Sim4_DIO}
  \end{subfigure}
  \begin{subfigure}[b]{0.4\linewidth}
    \includegraphics[width=\linewidth]{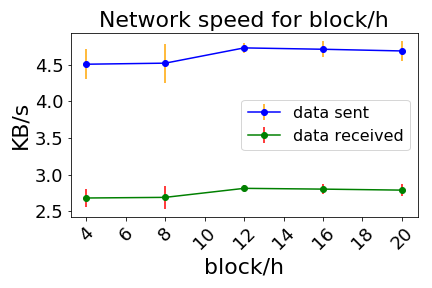}
        \caption{Network speed}
     \label{fig:Sim4_NetS}
  \end{subfigure}
  \caption{Fourth experiment results: increasing block generation speed}
  \label{fig:Sim4}
\end{figure}

Block generation speed does not affect neither memory nor disk usage (Figs.~\ref{fig:Sim4_Mem} and \ref{fig:Sim4_Disk}). Both metrics are stable, and show the same values than in the previous experiment.

Network speed shows a slight increase as block generation speed is incremented (Figure~\ref{fig:Sim4_NetS}).
Comparing this plot with Figure~\ref{fig:Sim2_NetS} (from Exp.~3 with $7$ tx/s), we can observe that the speed is the same, so the block generation speed does not significantly affect the network speed. The reason of these results is that from Bitcoin Core v0.13.0 onward, bitcoind implements compact block relaying~\cite{bip152}, a protocol for block relaying that does not re-send transactions in a block whenever the peer does already have these transactions.

Disk I/O speed shows a slight increase with block speed (Figure~\ref{fig:Sim4_DIO}), since more block metadata has to be written. 

\section{Conclusion and future work}
\label{sec:conclusions}

The nature of Bitcoin hinders its evaluation since, for security reasons, some parameters are only known at a local level. Hence, broader analysis cannot be performed without controlling all elements of the system. For instance, the network discovery mechanisms used for nodes to join the P2P network should provide a highly connected network with a difficult to estimate topology. The correctness of such network discovery mechanism cannot be properly addressed without knowing the exact topology of the produced network. The work proposed in this paper goes towards providing new tools for a global analysis of Bitcoin that would allow to better measure its performance and security.

This paper presents \textit{Kriptosare.gen}, a testbed based on Docker that allows deploying a Bitcoin network in order to emulate the real network behavior within a controlled environment. This testbed gives researchers full control over the generated network.
We have also presented and analysis of the cost of running simulation, taking into account different computer resources. Our results show that RAM and Disk usage are a limiting factor during the network creation phase. The experiments also show that network traffic increases with transaction generation speed in a linear fashion, and that transaction speed also affects CPU usage. Additionally, our results show that in a configuration similar to the real Bitcoin system (6 blocks/h) and with a throughput of 7 tx/s, an instantiation of 100 nodes can be run in a server with 4 CPU, 32GB RAM and 100GB of disk memory.

As future work, we intend to deploy the testbed over multiple servers in order to introduce routing, latency, and packet losses problems, and we will also simulate user behaviour and improve algorithms to detect undesired actions.
The testbed presented in this paper is a starting point for a multi-currency simulator, so the final idea is to add new private blockchains based on other cryptocurrencies like Zcash and Monero. 

\section*{ACKNOWLEDGMENT}

This work was partially funded by the European Commission through H2020 program, of which it is part ``TITANIUM" project (grant agreement No 740558). C. P\'erez-Sol\`a is currently affiliated to UOC, but part of this work was done while working at Universitat Aut\`onoma de Barcelona and Universitat Rovira i Virgili.

\addtolength{\textheight}{-12cm}  
                                  
\bibliographystyle{splncs04}
\bibliography{main}

\end{document}